# The role physics can play in a multi-disciplinary curriculum for non-physics scientists and engineers.

Edward F. Redish, Vashti Sawtelle, & Chandra Turpen
Department of Physics, University of Maryland, USA
For correspondence: redish@umd.edu

**Abstract**

At many physics departments a significant fraction of teaching is in support of engineers and scientists in other majors. These service courses are thus an automatic crucible of interdisciplinary interaction, and at times, strife. For example, the traditional algebra-based physics course is often considered by both biology faculty and students as having little relevance to their discipline. To address this issue, our multi-disciplinary multi-university team has been negotiating the role of a physics in the curriculum of life-science students; In *NEXUS/Physics* we have designed a class that stresses traditional physics skills but in contexts chosen to better meet the needs of life science students. Non-standard topics include chemical energy, diffusion and random motion, and thermodynamics with careful discussions of entropy, enthalpy, and Gibbs free energy. Explorations into how physics intertwines with an engineer's curriculum suggests places where analogous negotiations could lead to substantial modifications of physics courses for engineers that substantially enhance their value for engineering students.

**1. Introduction**

Reforming a curriculum can be quite a challenge. Instructional environments often evolve into stable patterns. Instructors know what to do, students know how to respond, grades are consistent with student expectations, and everything seems fine -- until some researcher starts asking new questions and we all learn that we have been fooling ourselves about the level of learning achieved. Part of the problem is that instructors tend to be subject experts. When they say something to their students or prepare a lesson, they bring meaning -- clusters of complex associations and related knowledge (Brookes and Etkina 2009; Redish and Gupta, 2009). When novice students do not possess these associations, what they mean by a response can be something quite different from what an expert instructor hears. This is why educational research teaches us that instructors need to have a good understanding not only their subject, but of their students -- what their students are bringing into the classroom and how they are thinking and learning. (Shulman 1986; Redish 1994)

The challenge of understanding our students becomes even more serious when the students we teach belong to a different disciplinary tradition than our own. Mathematicians teach physicists, physicists teach engineers, chemists teach biologists, and PhD professionals prepare school teachers. All have the problem not only of having to empathise with their students' inexperience about the instructor's subject, but with their students' experiences -- their knowledge, epistemologies, and self-identity stances coming from those students' identification with their own discipline. Gaps in an instructor's understanding of their students' disciplinary knowledge and stances can cause just as much trouble as an instructor's failure to understand student misconceptions.[1]

Over the past four years, a group of scientists, curriculum developers, and educational researchers have been creating a new introductory physics course for life science majors as part of a project supported by the Howard Hughes Medical Institute (HHMI)[2] and the US National Science Foundation (NSF)[3]: NEXUS/Physics. We believe that our experience in this project serves as a valuable case study for illustrating the point in the above paragraph. In section 2 of this paper, we describe some of what we have learned in the project about teaching physics to life science majors. In section 3 we speculate on how some of the lessons learned might play out in reforming introductory physics for engineers, and in section 4, we propose some general results that might be valuable for anyone reforming a curriculum for a reasonably well-defined population.

---

[1] When we use the term "misconception" we mean it in as a dynamic but predictable student misunderstanding, not as something necessarily part of a student's "naive theory". See Redish (2014).
[2] HHMI Grant: National Experiment in Undergraduate Science Education (NEXUS)
[3] NSF TUES Grant: Creating a Common Thermodynamics

## 2. Creating NEXUS/Physics

*2.1 Context:*

For many years, the number of biology majors and pre-health-care students taking introductory physics in US colleges and universities has been growing. At present, biology has the largest number of students of any major at the University of Maryland (UMd). All these students are required to take two semesters of physics, as are pre-medical students (whether they are biology majors or not). We refer to biology and pre-medical students collective as *life-science students*. The number of life-science students taking physics at UMd is substantial -- about 600 per year -- almost as many as the number of engineering students taking physics.

Both engineers and life-science students at UMd take physics in courses taught by faculty in our physics department. While we have traditionally taught our large number of engineers in a separate (calculus-based) physics course, life-science students have typically taken a catch-all (algebra-based) physics course that also served architects, computer science students, and any other student who needed a laboratory-based science class to satisfy a university requirement.[4]

In addition to the growth of the number of life-science students, there has been a growing demand from the communities of professional biologists, medical schools, and medical researchers for improvement in their students' undergraduate education.(National Research Council 2003, 2009; AAMC/HHMI 2009; AAAS 2011) This encouraged us to re-think what an appropriate physics course designed specifically for life-science students might look like.

*2.1.1 The goals of the project: Serving biology students better*

Since the algebra-based physics course typically taken by life-science students in the US does not include much that biologists find of direct value, physics is rarely a prerequisite for any upper division biology class. As a result, throughout the US a typical pattern has emerged: biology and pre-medical students take physics, but delay it until their third or fourth (final) year of university.

Physics faculty often view introductory physics classes as "non-negotiable". They see a coherent physics canon that might admit a few minor tweaks or change a few traditional examples for biological ones, but feel that all parts of the canon are essential for an "honorable" physics course. We took a different view. We noted that much of the content and approach of traditional introductory physics courses made a number of (often) tacit but unjustified assumptions:

- Physics should be taught in an essentially historic fashion, building up basic ideas from simple doable experiments and following by mathematical elaboration.
- Physics at the atomic and molecular scale should be saved for later courses.
- Much of the content of the class should be learning to reason mathematically with a focus on solving problems that are largely mathematical manipulation.
- The critical elements that must be preserved are those that are appropriate for a beginning mechanical, civil, or electrical engineer. (This controls the approach to mechanics, thermodynamics, electric currents, ....)

We decided to learn more about the biology curriculum and try to understand what value physics might have in a life-science students' curriculum, paying particular attention to physics might interact effectively with courses in biology, chemistry, and mathematics.

*2.1.2 Creating NEXUS/Physics: The process*

In order to understand how a physics class might better serve a life-science student, we built on what we had learned from research on how students in biology classes viewed occasional references to physics, and how biology students in physics classes interpreted occasional biology examples. (Watkins et al. 2012) We then spent the first year of our project negotiating the content and the approach with a team of physicists, biologists, education researchers, and biophysicists, with occasional meetings with chemists and mathematicians. We learned how our life science students interacted with physics, what physics might be seen as useful by biology faculty, and what biology might be useful (and interesting) to bring into a physics class.

---

[4] It is common in US colleges and universities that scientists are required to take courses in literature and the arts, and humanists are required to take laboratory science courses.

*2.2 What we learned: From negotiating with biology faculty*

We learned two rather surprising things from our negotiations with biologists:
- Biologists tend to take a dramatically different epistemological stance from physicists as to what is appropriate knowledge to introduce into introductory classes in their disciplines.
- The potential applications of physics to the life-sciences are potentially immensely broad -- too broad to deal with in a one year physics course.
- The biology curriculum had many topics where physics could provide significant support and coordination but these topics were traditionally not covered in algebra-based physics.

*2.2.1 Biologists and physicists often take differing epistemological stances in their Introductory classes.*

Here are some characteristic epistemological stances taken in almost every introductory physics class.(Redish and Cooke, 2013)
- P1. Physics stresses *reasoning from a few fundamental* (usually mathematically formulated) *principles.*
- P2. Physics stresses building a complete understanding of the *simplest possible (often highly abstract) examples* – "toy models" – and often don't go beyond them at the introductory level.
- P3. Physics emphasizes the ability to *quantify* a view of the physical world, *model with math*, and *think with equations*, both qualitatively and quantitatively.
- P4. Physics classes are concerned with *constraints* that hold regardless of internal details (e.g. conservation laws, center of mass).

These elements will be familiar to anyone who has ever taught (or taken) introductory physics. However, we rarely articulate them for students – and <u>none</u> of these elements are typically perceived by students as present in introductory biology. In their introductory classes, biologists have other concerns. Here are some epistemological stances communicated to students in introductory biology classes.
- B1. *Biology is highly complex*. Biological processes commonly involve interactions of many component parts, leading to emergent phenomena (including the "property" of life itself).
- B2. Introductory *biology rarely emphasizes quantitative reasoning* and problem solving.
- B3. Biology is subject to *historical constraint* in that natural selection can only act on pre-existing molecules, cells, and organisms. Much therefore depends on specific properties of real cases.
- B4. Introductory biology is largely *descriptive* since there are many terms, organisms, and systems that students need to be familiar with.
- B5. Introductory biology focuses on *real examples* and *structure-function relationships*.

These differences have powerful implications for life-science students in a physics class. The students are highly suspicious when physicists introduce toy models (the spherical cow in a frictionless vacuum) and then turn many mathematical gears. They infer that when the systems are simplified, structure is changed too much, so function is suspect; and therefore that the mathematical games physicists play are not likely to not really be relevant for them. (Watkins et al. 2012)

This suggests that in our physics for life-science majors classes,
- We need to be much more explicit about the motivation for our simple models and what they teach us;
- We need to include biological examples that the students perceive as *authentic* -- useful and relevant to their understanding of biology.[5]

---

[5] Of course "physics is a liberal art" and every educated person should know the physics that we traditionally teach, but if the relevance of physics is remote, it is hard to make a case for maintaining space for it in a life-science curriculum that is increasingly tight as the breadth of biological topics of direct value to a life-scientist grows.

*2.2.2 "Physics for life scientists" is too broad to provide much guidance.*

One of the first things physicists often do when trying to design a physics for life scientists is to look for applications of physics to the life sciences at the professional level to see what the students might need in the future. This doesn't work for two reasons.

First, the field of life science is too broad. It includes microbiologists, medical professionals, specialists in sport medicine, neuroscientists, physiologists, ecologists, and evolutionary biologists. Cases can be made for including every topic that could conceivably be taught in an introductory physics class. Kinesiologists need to understand projectile motion and torque; botanists need to understand surface tension, capillary action, and negative pressure; neuroscientists need to understand circuit theory. And if we take the stance that any professional scientist who uses a complex technological tool ought to have some idea of how it works, then we need to include discussions of x-ray machines -- and therefore photons and the photoelectric effect, and MRI machines -- and therefore magnetic interactions and gyroscopes.

Second, it is a superficial approach, thinking of physics as a set of facts to be learned rather than a way of thinking to be mastered. The course that would come out of such considerations is far too broad and far too shallow to provide any student with the kind of epistemological value that physics is well designed to support: learning how to reason from principles, learning how to model and reason with mathematics, learning how to build coherence around consideration of mechanism, and much more. (Redish and Hammer, 2009)

A major goal of our re-design was to make sure that life science students had the opportunity to feel how it felt to understand something in physics and to learn how physics could be useful for their work in biology. To reach this goal we rethought not only the content and the placing of the physics course in the program of a life science student but also
1. What does it look like to bring physics and biology into meaningful interactions in specific reasoning tasks?
2. How and in what ways can we leverage students' chemical and biological expertise in reasoning about physics?
3. Which phenomena or reasoning contexts highlight the value of multiple disciplinary perspectives?

Answering these questions required partnerships between biologists and physicists, where the knowledge, problems of interest, and assumptions of both disciplines are valued.

In our discussions with biologists during the first year, in our interviews with the students, and in our review of the biology and chemistry education research literature, we learned that there were topics where the epistemological approach traditionally taken by introductory physics students could be extremely valuable and coordinate well with their other science classes.

*2.3 Repurposing the class to fit the needs of the students*

Since there were so many possibilities in a course for life science students, we decided to focus on a few broad goals:
- To serve common needs of the broad population of students in life science.
- To help life science students develop the strengths normally associated with physics:
  - Building an understanding of physical mechanism in a way that allows them to create a sense of coherence and sense-making.
  - Appreciating the value of mathematical reasoning in qualitative analysis as well as calculational situations.

To do this, we made a number of explicit decisions. (Redish et al., 2014)
1. We chose to design the class so that it focuses on classes required of <u>all</u> life sciences students -- cellular biology, evolution, biochemistry, and general chemistry. To make physics mesh better with the curricula of students taking these classes,

a. We envisioned it as a second year class and therefore required two terms of introductory biology, one term of general chemistry, and two terms of calculus.[6]
   b. We included topics that we viewed as critical for life-science students where a physics approach could help: chemical energy and bonding (Dreyfus et al. 2014), random motion including diffusion (Moore et al. 2014), entropy and Gibbs Free Energy (Geller et al. 2014), motion of an in fluids, and electrostatics in fluids.
   c. In order to make room for the new topics we reduced or eliminated traditional topics that have less general relevance to the life-sciences: projectile motion and inclined planes, circular motion, universal gravitation, magnetism, special relativity.
   d. We emphasized leveraging what the students know from their other classes to support their engagement in reasoning about physics.
2. We chose to create examples and lessons that were perceived by the students as having biological authenticity. Building from things that the students had learned in biology and chemistry, we created physics lessons in biological contexts that resulted in the students feeling that they were developing a better and more coherent understanding of topics they felt were important. (Svoboda et al. 2013)
3. We chose to build on extensive education research on epistemology development in order to develop student competencies, not just knowledge.

We decided that we would consider our class successful if
- Life-science students see that adding physics into their reasoning about a biological phenomena adds something to their thinking, or (even better) that they readily take up physics ways of thinking to reason in places that they previously did not.
- Life-science faculty teaching upper division classes decide that physics was valuable enough to them that they make it a prerequisite for upper division classes.

In the second and third year of the project (2011-13) we developed materials and taught the class to small numbers of students (N ~ 20). We created a Wikibook of online webpages, homework problems we deemed relevant (after negotiation with biologists), group learning lessons that focus on important biological examples, and laboratories that study random motion, motion in fluids, electrostatics in fluids, and spectroscopy using authentic biological tools. Many of the preliminary materials for this course are available on the web at http://nexusphysics.umd.edu.

In these two years the course was taught with students reading materials before class and the lecture classes including extensive discussion. We videotaped all lectures, group learning lessons, and we interviewed students throughout the year to carry out case studies. We did pre- and post-testing with a variety of existing instruments such as the FMCE (Thornton and Sokoloff 1998), the MPEX (Redish et al. 1998), a lab survey from Etkina & Murtha (2006) and some new ones of our own. We learned a lot from our research with students, and we think we can draw some general conclusions.

*2.4 What we learned: From watching and listening to students*

We learned much from our careful interaction with students, both inside of class and in our interviews. We think that some of these lessons have general applicability to classes taught to populations with different disciplinary expectations. Here are three.
- Many of our standard examples need to be rethought.
- We need to understand what in our curriculum students might potentially value.
- We need to understand and respect students' disciplinary perspectives.

*2.4.1 We need to reframe our standard examples: "Between the sheets"*

Two of the broad physics competencies that we want students to take away from our class are:

---
[6] At UMd we have a special calculus class for biologists that in addition to covering the basic elements of calculus also focuses on exponentials and probability. These latter are perhaps more important than much calculus, though calculus concepts are referred to throughout.

> *Physics is about something real. Whenever you think about a physics example, start from a mental image of a physical situation and refer everything back to it.*

In physics our equations, our graphs, our diagrams, are all about some physical situation and are intended to inform us about some aspect or relationship of that situation, building a richer and more multi-dimensional view of it. This element of referring every representation back to a single vision is what's missing for many students in introductory physics. They want to "answer-make" rather than "sense-make" and fail to build that underlying physical picture. Having a physical model can (and should) guide them in understanding the mechanism of what's happening, in deciding what they have to pay attention to and what they can ignore, and in figuring out what principles are relevant with what restrictions.

Second, we explicitly tell our students why we do "toy models":

> *The "style" of physics is to simplify. We always try to find the simplest example that illustrates a principle so we can understand it fully. We then use that example as the core to elaborate our thinking into more realistic situations.*

This is our way of getting a foothold that we can make sense of, in order to imbed understanding into organizing and "finding the physics" in a complex situation.

In the first year we taught NEXUS/Physics, we had a dramatic example of how there is a dynamic tension between these two aspects of physics thinking. We were studying electric fields and potential. A standard example is the "infinite flat sheet of uniform charge." This is a nice example since the math simplifies dramatically. Because of Coulomb's law, any electric field has to look like a charge divided by the square of a distance (times a universal constant chosen to set the measurement scale). When we have an infinite sheet, we have no "charge" we can use – it's infinite – we only have the charge per unit area. This already has units of charge divided by length squared, so there is no room for any other distance in a formula for the electric field. The result is that the field has to be a constant, independent of the distance from the sheet.

This seems strange, but it actually makes sense. To get the total effect of the infinite sheet you have to add up the Coulomb's law contributions from each of the bits of charge in the sheet to the field at the point you are sitting at. Each bit of charge contributes a field vector that points along the line to your point from the charge that is proportional to one over the distance to that charge squared. As you go farther away to the more remote charges, they contribute less and less. Also, each distance charge is paired with another distant charge equally far away on the other side and these contributions tend to cancel – and cancel more and more the farther away you get. The result is that for the entire infinite sheet, if you are a distance $s$ from the sheet, only the circle right beneath you of radius about $10s$ contributes significantly to the field you detect. So although we say we have "an infinite sheet" that's not what we mean. We mean: we have a flat sheet and the edges are far enough away that we don't have to worry about them. So the result is: only a circle of radius $10s$ matters. As you increase $s$, the effect of each charge on the sheet falls off like $1/s^2$, but the amount of charge you see grows like $s^2$. These two effects cancel to give a constant field.

The result of having a constant E field simplifies a lot of the math. The potential that goes with a constant field is just linear (since the derivative of the potential is the E field) so the math is really simple - school algebra. All those complex "curvy $1/r^2$" functions and vector integrals add up to give straight lines. It looks just like the same math for flat-earth gravity – where we take the gravitational field to be constant always pointing in the same direction.

So here's where we ran into something interesting. Students read our webpage on fields near a sheet (and it was read and summarized by 20 out of 20 of the students). Then in class we asked the following clicker questions:

> *If two uniform sheets of equal and opposite charge can be treated as if they were infinitely large, which of the following graphs might serve as a graph of (A) the x-component of the electric field and (B) the electrostatic potential as a function of the coordinate x along the dotted line?*

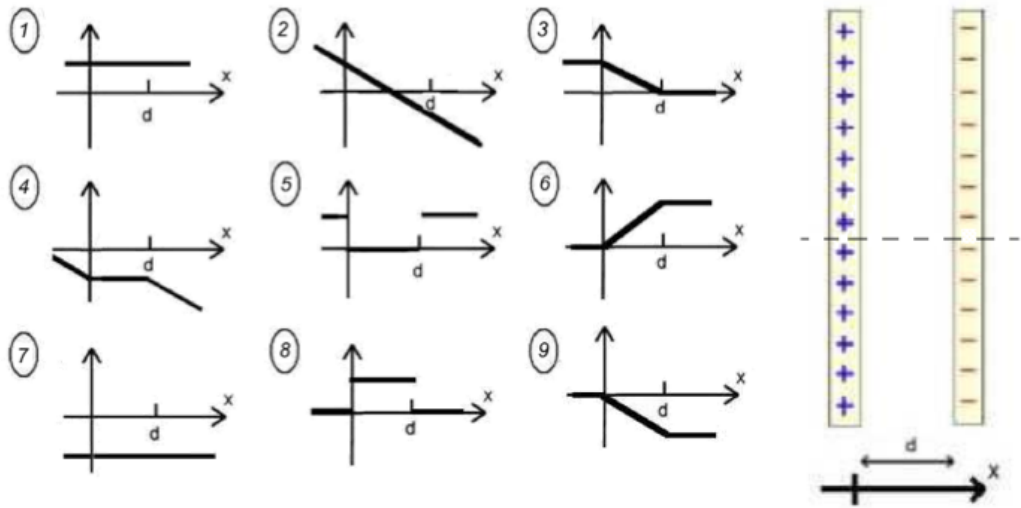

Figure 1: A traditional toy model problem

Since the fields from each sheet are constant and since the charges are equal and opposite, outside of the sheet the two fields cancel, and between them they add. The result is that the E field (x component) looks like graph 8. Looking for graphs whose (negative) derivative looks like 8, we see that both 3 and 9 work for the potential. That's OK since for the electric potential we can choose any reference point as zero potential (as we can for height when we are talking about gravity). It's only the shape that matters physically.

On the first question – what does the electric field look like – the results were gratifying. Of the 19 students present, 16 chose answer 8. But one student complained. He said, "I didn't like any of them." When asked why, he responded, "Because when you get near to the sheets you'll see the individual charges and the field has to go to infinity." The lecturer brushed him off with a brief comment about resolution – that it would only happen really, really close and we wouldn't see it on this scale and anyway we were ignoring individual charges in this simple model.

On the second question – what does the potential look like – the results were more mixed. About half chose answer 3, but the rest were all over the lot. But one student (a different one) said, "It can't be any of them." When asked why, he responded, "Because if you are sitting exactly between the plates the result has to be zero. If you are sitting there, for every positive charge on one sheet there is a negative charge on the other sheet an equal distance away that will cancel. Also, it has to eventually asymptote to zero for large values."

This was absolutely top quality physics reasoning. He was using a physical picture and using it with a correct symmetry argument – another strong tool in the quiver of good physics thinking. (This was something we had been trying to model in the class, but not fussing too much about.) Finally, he was focusing on limiting cases, another standard tool we try to get our students to use.

Although at first we were pleased, on second thought a problem appeared. We seemed to want them to be looking at this model example NOT in terms of the basic physical elements but as a toy model that suppressed the underlying physical picture. Since a major goal of the class is to teach my students to *seek consistency*, how should we present this model?

The lecturer tried to be encouraging and praise the students' thinking but still bring the class back to being able to using the simplified model while "suspending our disbelief" – using it even though we know that "when we go to infinity the sheets that look infinite when we are close will turn into looking like point charges when we are far away", and "we'll ignore the fact that the charges are actually quantized and treat it as if it were perfectly smooth since we don't see the individual charges until we are really close and then we'll just 'close our eyes' for a moment while we are passing through the sheet".

Really what it's about is modeling. (Hestenes 1992) When we have complex situations – like lots and lots of charges – it's great to have "a stake in the ground"; an example where we know an "exact" answer that we can refer to that serves as a starting point for further examples and elaborations. But in introductory physics for physics majors, we tend to focus on the simplicity of the models and not on the complexity of thinking about where they come from, what their validity is, and what value we can make of them. Thanks to the epistemological stances they have been exposed to (especially B1 and B5) life science majors are often uncomfortable with toy models in which the structure of the system is dramatically changed and the value of understanding the simplified system is not made clear.

Many of our canonical simplified examples are focused on creating situations where the math simplifies. But pretending that the physics is simple by hiding its deep structure, both ontological and epistemological (i.e., what is it we are actually talking about and how is it we decide we know), seems unfair to our students and not the best way to start them on learning physics. It seems to be a particular problem with life-science students, many of whom seem particularly unwilling to "suspend their disbelief" about an unrealistic situation.

*2.4.2 We need to help students see physics as relevant to things they care about: Violet doesn't hate physics anymore*

It is common for physics faculty to have some starting assumptions about students entering our classrooms. Some of these are useful in terms of the kinds of everyday experiences students will have encountered in the real world (i.e. they have driven a car or kicked a ball); others are assumptions we have about how students will feel about a physics class. When we think about teaching life-science students introductory physics, we commonly fall back on assumptions about these students not being very good at maths, being afraid of physics and the grade they might receive, and being primarily interested in particular kinds of examples (e.g. health applications). In rethinking our physics for life science majors course, and really listening to students we have realized that some of these starting assumptions are unfounded and potentially unproductive for impacting the ways students experience physics.

An example that challenged our starting assumptions comes from a longitudinal case study we developed around one student, "Violet", in the second iteration of the NEXUS/Physics course. In the early days of the course, Violet seemed to fit many of these starting assumptions about life-science majors. On the first day of class, Violet publicly voiced her expectation that physics would "…make my life a living hell…as a biologist I've avoided physics for the longest time." In an interview she reiterated this perspective saying,

> *I feel like I came in with a lot of hatred towards physics in general….I never took [physics] while I was in high school, so coming in and knowing for the [biological sciences] curriculum that I had to take two semesters of physics, something that I never had any exposure to, I wasn't very happy about. And I'm more of a bio and chemistry person, and I just take physics for granted. I was like, oh it falls. OK. Sure.*

From these starting stances it might seem that Violet is a lost cause. She expects physics to be difficult, describes it with hatred, and positions it as in contrast to herself as a "bio and chemistry person." However, in observing Violet in classroom situations and following her in interviews we see that Violet's stance was in part a product of what she thought physics was about. ("Oh it falls. OK. Sure.")

The NEXUS/Physics course provided opportunities to dynamically interact with this stance, and ultimately changed Violet's relationship with physics.

Throughout the NEXUS/Physics course, we created opportunities for physics to interact with students' biology knowledge. For example, Violet described one group problem-solving activity that asked students to consider the impact of modeling a DNA molecule as a spring. Reflecting on this task in an interview, Violet recounts how she helped explain the structure of DNA to the teaching assistant (TA).

> *And I went up to the chalkboard and like started drawing a string of DNA to show [the TA] because she was like I don't know any biology. And I was like ok I'll help you with this part, and I mean it just clicked so much more easier for me. I was like well DNA is blah, blah, blah there's this, this, and this. Maybe this factors into this and it takes more force for this to happen.*

In this quote we see how Violet's identification with and sense of competence in biology became an important part of how she interacted with this physics class. During the class, we saw multiple episodes of Violet bringing biology into her physics reasoning, with increasing confidence. By the end of the course Violet describes having a view about physics that was more connected with her knowledge from chemistry and biology,

> *Yeah, I feel like if I were in a regular [physics] class I wouldn't like the physics because it just doesn't relate. Because me as a biologist, I like biology and chemistry so much more than I did physics. But now that I can see the relationship between all three, it's kind of made me like physics more.*

So what did we learn from studying how Violet's relationship with physics developed over the NEXUS/Physics course? We recognized that our students have many more experiences to draw from than simply the everyday experiences like driving a car. These students commonly have a rich set of knowledge from biology and chemistry that may be productive resources for reasoning about physics. We also learned that providing students opportunities to draw on their outside expertise can productively interact with the starting stances that students enter our classrooms with. Violet entered our NEXUS/Physics course expecting to hate physics, and thinking of physics as dealing with things like balls falling and roller coasters going down hills. By creating opportunities to leverage her outside expertise she left the course with a more positive orientation toward physics, and describing the subject as productively linked to biology and chemistry.

We suggest that Violet's positive affect[7] and strong identification with biology were productive levers for influencing her relationship with physics. As Violet had opportunities to see physics as useful for understanding ideas in biology that she cared about, the domains of physics and biology became linked in important ways. Linking these domains resulted in reducing tensions at times (such as seeing "being a bio person" as less in contrast with success in physics). Importantly, in Violet's outgoing relationships with physics and biology we see Violet identifying places where physics and biology "coincide" or "come together" and associating these moments with feeling glad or happy. Recruiting Violet's relationship with biology into experiences of learning physics impacted her relationship with physics as well as her sense of how physics and biology are linked.

Our work suggests that in designing learning environments we should consider carefully the identities and affective orientations students bring into our classrooms, particularly at the advanced levels of students' academic trajectories. The case of Violet provides insight into how a student can feel expert in a classroom without feeling like an expert in the discipline. Ultimately, this caused us to rethink the starting assumptions we have about life-science students in our physics classrooms. In our redesign we have begun to think about how to draw on these student trajectories and acknowledge the wealth of knowledge and expertise life-science students bring into our classrooms, and how we can draw out those experiences in productive ways to impact participation in physics.

*2.4.3 We need to understand and respect disciplinary perspectives: Gregor*

When we first began to integrate chemical energy into our class we had a dramatic realization about how different disciplinary perspectives and assumptions played a role. (Dreyfus et al. 2014)

From our discussions with biologists and chemists, and from our exploration of the chemistry education literature, we decided that one place where physics instruction might be able to play a role in helping life-science majors is in their understanding of chemical bonding and exothermic reactions. By this we don't mean a mathematical and quantum mechanical treatment of bonding. Rather, we learned that introductory biology and chemistry teach a little about chemical bonds, but basically give heuristic rules rather than creating a coherent and mechanistic picture. For example:

1. You need to put in energy in order to break a chemical bond (General Chemistry I)
2. The molecule ATP[8] is the "energy currency" of the cell and serves to provide energy for a variety of essential biological processes. (Introductory Biology I).

---

[7] By "affect" we refer to the cluster of factors that include emotional response, motivation, self-perception, identity, etc.
[8] ATP is adenosine triphosphate. The removal of phosphorus from ATP by reaction with water to produce ADP (adenosine diphosphate) is an exothermic reaction that releases energy. Depending on the

Both of these statements are correct, but many biology students have a hard time making sense of it. We often heard what may be called the "piñata model" in which putting a little energy into a bond allows it to release a lot of "pent up" energy. One student expressed this as follows:

> *But like the way that I was thinking of it, I don't know why, but whenever chemistry taught us like exothermic, endothermic, like what she said, I always imagined like the breaking of the bonds has like these little [energy] molecules that float out, but like I know it's wrong. But that's just how I pictured it from the beginning.*

This is identified in the chemistry education literature as a persistent misconception. (Galley 2004) We expected that a careful physics treatment of the energy in the reaction might help. But the first time we gave one of the standard questions (from Galley) we wound up in an interesting discussion. Some of the students did indeed seem to be confused about where the energy was coming from. But others expressed a more sophisticated view. Gregor responded (in an interview)

> *I guess that's the difference between like how a biologist is trained to think, in like a larger context and how physicists just focus on sort of one little thing. Whereas like, so I answered that it releases energy, but it releases energy because when an interaction with other molecules, like water, primarily, and then it creates like an inorganic phosphate molecule that has a lot of resonance…. So like, in the end releases a lot of energy, but it does require like a really small input of energy to break that bond.*

What Gregor is essentially pointing out is that if we think of the ATP + $H_2O$ reaction as taking place in isolation, then it is clear that the source of the energy release is in the fact that a weak bond is broken and a strong one formed. The system goes from a small negative potential energy to a more negative one, releasing energy. But if one thinks about ATP in the context of a biological system, then water is always available and one doesn't need to mention it. You can look at it that having some ATP gives you the opportunity to carry out processes that require energy. You don't have to think about how it does it any more than you think about how having gas in your car lets you drive. You just need to know that you need it.

This is in fact similar to the way we think about energy in everyday speech. We consider that glucose and other foods "contain" energy and specify the amount on our packaging in Joules or Calories. In fact the energy is released only when the food is combined with oxygen and the release of energy comes from making the strong $C + O_2 \rightarrow CO_2$ bond. But since we assume that oxygen is always available we don't typically mention it.[9]

What we take from this example is that we were taking too narrow a view when we viewed "the energy stored in a chemical bond" as a misconception. Rather, it is perfectly reasonable to use that language -- in a context in which the needed reactants are always available. But we expect students to be able to unpack how that happens if needed, and to feel comfortable that they can reconcile the two statements given at the beginning of this section by analyzing the physical mechanism involved.

This example gave us a new goal: interdisciplinary reconciliation. Not only did we want to help our students understand this example from a physics (or chemistry) point of view that isolated the reactants, but to make sense of how it was talked about in biology and to understand the contexts in which each approach was relevant. We had to develop an understanding of and a respect for the perspective of biologists.

Treating what our life-science students bring into our classes with respect requires that physicists do much more than throw in a few biological examples.

---

chain of processes intervening in ATP + $H_2O \rightarrow$ ADP + $P_i$ energy is delivered to accomplish various biological ends.

[9] We are grateful to Leslie Atkins Elliott for pointing out this analogy.

- It requires deep thinking about the character of scientific inquiry in the two professions.
- It requires deep thinking about the nature of thinking and learning in science.

We learn not just about teaching, but about the way the two disciplines think about and approach scientific problems.

## 3. Thinking about physics for engineers: Some speculations

What we have seen in our development of NEXUS/Physics for life-science students is that the profession of biology has a somewhat different epistemological orientation from that of physics, especially when considering instruction in introductory classes. In addition, life-science students bring in expectations associated with their identification as biologists (or pre-medical students) that may create a mismatch with the expectations of traditional physics instructors and the environment of traditional physics instruction. We found that understanding what the students were bringing to the class provided us powerful opportunities for both improving their engagement with and learning of physics. This knowledge was not just what we might view as "misconceptions", but what we might view as <u>unacknowledged strengths</u>: their disciplinary knowledge from other courses, their understandings of the nature and goals of their discipline, and their interest and motivation. A natural next question is: This isn't just about biology students, is it?

While our research group has not recently carried out a major curriculum reform project in physics for engineering students, we have been carrying out basic research with engineers in physics and their follow-up classes. Some of us have been teaching introductory design for engineers and physics for engineers. From our interactions with engineers and our observations with engineering students, (Gupta and Elby, 2011; Hull et al., 2013; Danielak et al., 2014) we speculate that there are similar disciplinary issues that potentially create a mismatch between the expectations of physics instructors and (at least some of) their engineering students.

In analogy to our list of epistemological stances taken about introductory science classes by physicists and biologist in section 2.2.1, a first draft of such a list for engineers and engineering students might include the following elements.

E1. Engineering often prioritizes knowing/learning for the purpose of *design* rather than for creating *causal models* of natural phenomenon.
E2. Although engineering students expect to use mathematics in their science classes, many expect to use it primarily for *calculation* of specific examples rather than for *sense making* with symbolic representations.
E3. Equations in physics classes are expected to give deterministic answers (calculating a force, an acceleration, etc.) and are the end point of problems. In engineering, the products of equations are often starting points for iterations that lead to optimizing the solution to fit the non-idealized aspects of real world environments.
E4. While knowing in an intro-physics class might privilege knowledge generated from idealized equations and situations, finding engineering solutions might require relying also on practical knowledge, kinesthetic knowing, rules of thumb, etc.

Each of these has implications for how we might most effectively interact with our engineering students in physics. Engineering students often prefer *realistic* examples and, like the biology students, might not see the value of our toy model examples if that value is not explicitly discussed. In physics problems, we might ignore friction, or air drag, consider point objects etc. But including these is important when designing something in the real world where dissipation and noise cannot be ignored. In that sense, engineering is more firmly situated in the physical world than an abstracted idealization of it. The strong emphasis on abstractions in traditional introductory physics classes for engineers can lead to a mismatch of expectations, making the 'regular' intro-physics problems seem not only irrelevant, it can generate a sense that idealized models are not relevant when solving problems in the real world. (Gupta et al. 2014)

The challenge becomes to design a physics course that can blend the productivity of modeling tools common to introductory physics with the full complexity of real-world phenomenon. Here, as in our life-sciences case study, we expect that understanding and relying on the students' strengths -- experience with the real world, knowledge of their discipline, and motivation for their profession -- might provide us with powerful levers for creating a course that works better for engineering students.

## 4. Conclusion: Understanding and respecting what your students bring to your class can change the way you see your teaching

The idea that students bring conceptual ideas about how the world works into our physics class is old and well documented many times over. (McClosky 1983; diSessa 1993) What is often not appreciated is that how much the way they use those concepts depends on their expectations about the class, their assumptions about the nature of the knowledge they are learning, the task, and how they interpret the context in which questions are asked. (Hammer et al. 2004; Redish 2014) When we are teaching a population of students in a discipline not our own, those expectations can be characteristic of their discipline and may well be unfamiliar to an instructor.

To summarize, we have learned two things from our case study example of NEXUS/Physics.
1. When we consider what it is our students bring into the class we should not only consider their weaknesses, but their strengths. In an interdisciplinary situation (instructor of one discipline, students of another) these may include the students' knowledge of their own discipline.
2. Some of these strengths include the students' affective responses -- their motivation, emotional responses, and sense of identity. If these are ignored, they can create barriers and hostility that inhibit learning. But if we understand and work with them, they can be levers for improving engagement and learning.

We may summarize our advice for reformers in interdisciplinary situations with heuristic organized around two ideas: _Respect_ and _Reflective Design_.
- _Respect_ your target discipline and understand what their goals are for their students in your class.
- _Respect_ the knowledge your students bring into the class. They will be making sense of what you teach them using that knowledge.
- _Respect_ your students as human learners. They are not robots to be programmed. Give them room to explore, play, and learn in a natural way.
- _Design_ your learning goals to meet the students where they are and to help take them where they need to go. This requires a good understanding of the track of their discipline.
- _Design_ your learning environments so as to give your students the opportunity to learn not only the content you expect them to learn, but also the learning methods that are appropriate for learning the conceptual, factual, and epistemological content of your discipline.

While we have learned these lessons while teaching majors in other disciplines, the process of reflective design is certainly valuable for teaching students in our own discipline as well. When teaching students in other disciplines, we may be unaware of what they bring into our class in terms of expectations, experience, needs with respect to their curriculum, and affective stances. Curriculum reforms for majors in other disciplines than our own are likely to be much improved by developing an understanding of these issues. But our own majors are novices being acculturated into a complex social system. Having a better understanding our own students along these lines is likely to be equally valuable.

**Acknowledgement**

We are very grateful to the entire NEXUS/Physics team including collaborators from many universities for their many contributions to the project. We acknowledge significant and valuable contributions to this manuscript from Ayush Gupta. This work was supported in part by a grant from the Howard Hughes Medical Institute and by grants from the US National Science Foundation.

**References**

AAAS (2011). *Vision and change in undergraduate biology education: A call to action*: AAAS Press.

AAMC/HHMI (2009). *Scientific Foundations for Future Physicians: Report of the AAMC-HHMI Committee*

Brookes, D. & Etkina, E., (2009). Force, ontology and language. *Physical Review, Special Topics, Physics Education Research*, **5,** 010110;

Danielak, B., Gupta, A., & Elby, A. (2014). The Marginalized Identities of Sense-makers: Reframing Engineering Student Retention. *Journal of Engineering Education* 103:1, 8-44;

diSessa, A. A. (1993). Toward an Epistemology of Physics, *Cognition and Instruction*, **10,** 105-225;

Dreyfus, B., Geller, B., Gouvea, J., Sawtelle, V., Turpen, C. & Redish, E. (2014). Chemical energy in an introductory physics course for the life sciences, *Am. J. Phys.* in press.

Etkina E. & Murthy, S. (2006). Design labs: Students' expectations and reality, *AIP Conf. Proc.*, **818** 97-100.

Galley, W. (2004). Exothermic bond breaking: A persistent misconception. *J. Chem. Educ.* **81** (2004) 523.

Geller, B., Dreyfus, B., Gouvea, J., Sawtelle, V., Turpen, C. & Redish, E. F. (May, 2014). Entropy and spontaneity in an introductory physics course for life science students, *Am. J. Phys.* in press.

Gupta, A., Danielak, B. & Elby, A. (2014) *Physical Review - Special Topics -PER*, under review

Gupta, A. & Elby, A. (2011). Beyond Epistemological Deficits: Dynamic Explanations of Engineering Students' Difficulties with Mathematical Sense-making. *Int, J. of Science Educ.,* **33**:18, 2463-2488.

Hammer, D., Elby, A., Scherr, R. & Redish, E. (2004). Resources, framing, and transfer, in *Transfer of Learning: Research and Perspectives*, J. Mestre (ed.): Information Age Publishing.

Hestenes, D. (1992). Modeling games in the Newtonian World, *Am. J. Phys.* **60** 732-748

Hull, M., Kuo, E., Gupta, A. & Elby, A. (2013). Problematizing problem-solving rubrics: Enhancing assessments to include blended mathematical and physical reasoning throughout the solution, *Physical Review Special Topics - Physics Education Research*, 9:1, 010105;

McCloskey, M. (1983). Naïve theories of motion, in Dedre Gentner and Albert L. Stevens, Eds., *Mental Models*, 299-324: Erlbaum.

Moore, K., Giannini, J. & Losert, W., (2014) Toward better physics labs for future biologists, *Am. J. Phys.*, in press.

National Research Council (2003). *Bio 2010: transforming undergraduate education for future research biologists:* National Academy Press.

National Research Council (2009). *A New Biology for the 21st Century:* National Academy Press.

Redish, E. F. (1994). Implications of Cognitive Studies for Teaching Physics, *Am. J. Phys.*, **62**, 796-803.

Redish, E. F., Saul, J. M., & Steinberg, R.N. (1998). Student Expectations In Introductory Physics, *Am. J. Phys.* **66,** 212-224.

Redish, E. F. & Hammer, D. (2009). Reinventing college physics for biologists: Explicating an epistemological curriculum, *Am. J. Phys.*, **77**:7, 629-642.

Redish, E. F. & Gupta, A. (2010). Making Meaning with Math in Physics: A Semantic Analysis, *Physics Community and Cooperation, Vol. 1* GIREP Conf. Proc., Leicester, UK, August 20, 2009, 244-260.

Redish, E. F. & Cooke, T.J., (2013). Learning Each Other's Ropes: Negotiating interdisciplinary authenticity, *Cell Biology Education - Life Science Education*, **12,** 175-186.

Redish, E., Bauer, C., Carleton, K., Cooke, T., Cooper, M., Crouch, C., Dreyfus, B., Geller, B., Giannini, J.,Svoboda Gouvea, J., Klymkowsky, M., Losert, W., Moore, K., Presson, J., Sawtelle, V., Thompson, K., Turpen, C. & Zia, R. (May, 2014). NEXUS/Physics: An interdisciplinary repurposing of physics for biologists, *Am. J. Phys.* in press.

Redish, E., (2014). Oersted Lecture 2013: How should we think about how our students think? *Am. J. Phys.* in press.

Shulman, L. S. (1986). Those Who Understand: Knowledge growth in teaching, *Educational Researcher*, **15**:2 4-14.

Svoboda, J., Sawtelle, V., Geller, B., & Turpen, C. (June 3, 2013) A Framework for Analyzing Interdisciplinary Tasks: Implications for Student Learning and Curricular Design, *Cell Biology Education - Life Science Education* **12,** 187-205.

Thornton, R.K., and Sokoloff, D.R. (1998) Assessing student learning of Newton's laws: The Force and Motion Conceptual Evaluation, *Am. J. Phys.* **66** 228-351.

Watkins, J. & Elby, A. (June 3, 2013). Context dependence of students' views about the role of equations in understanding biology, *Cell Biology Education - Life Science Education* **12**, 274-286.

Watkins, J., Coffey, J. E., Redish, E. F. & Cooke, T. J. (2012). Disciplinary Authenticity: Enriching the reform of introductory physics courses for life science students, *Phys. Rev. ST Phys. Educ. Res.*, **8** 010112.